\documentclass[aps,floats,epsf, twocolumn,showpacs ]{revtex4}
 \usepackage{graphics}

\begin{document}

\title{Odd integer quantum Hall effect in graphene}

\author{Bitan Roy}

\affiliation{Department of Physics, Simon Fraser University,
 Burnaby, British Columbia, Canada V5A 1S6}

\begin{abstract}
A possible realization of Hall conductivity, quantized at odd integer factors of $e^2/h$ for graphene's 
honeycomb lattice is proposed. I argue that, in the presence of \emph{uniform} real and pseudo-magnetic 
fields, the valley degeneracy from the higher Landau levels can be removed. A pseudo-magnetic 
field may arise from bulging or stretching of the graphene flake. This may lead to observation of 
plateaus in the Hall conductivity at quantized values $f e^2/h$, with $f=\pm 3, \pm 5$ etc, which have 
not been observed in measurement of Hall conductivity. However, in a collection of noninteracting Dirac 
fermions living in the honeycomb lattice subject to real and pseudo field, the zeroth Landau level still enjoys 
the valley and the spin degeneracy. Upon including the Zeeman coupling, the spin degeneracy is removed 
from all the Landau levels. The effects of short ranged electron-electron interactions are also 
considered, particularly, the onsite Hubbard repulsion (U) and the nearest-neighbor Coulomb repulsion (V). Within 
the framework of the extended Hubbard model with only those two components of finite ranged Coulomb repulsion, it 
is shown that infinitesimally weak interactions can place the system in a gapped insulating phase by developing a 
\emph{ferrimegnatic} order, if $U>>V$. Therefore, one may expect to see the plateaus in the Hall conductivity at 
all the integer values, $f=0,\pm 1,\pm 2, \pm3, \cdots$. Scaling behavior of interaction induced gap at $f=1$ in 
presence of finite pseudo flux is also addressed. Qualitative discussion on finite size effects and behavior of the 
interaction induced gap when the restriction on uniformity of the fields are relaxed, is presented as well. Possible 
experimental set up that can test relevance of our theory has been proposed.
\end{abstract}


\maketitle

\vspace{10pt}

\section{Introduction}

Carbon atoms bonded in a two-dimensional honeycomb lattice, named as \emph{graphene} \cite{Novoselov}, 
recently engaged both theoretical and experimental attention due to a variety of unconventional properties. 
The lack of inversion symmetry in graphene lattice structure leaves the Brillouin zone with special points 
at its corners, allowing one to linearize the spectrum around those points, which leads to a conical 
dispersion. At filling one-half, when the chemical potential coincides with the apex of the cone, excitations can 
be described in terms of massless, chiral Dirac fermions, where the Fermi velocity $(v_F=C/300)$ plays the 
role of the velocity of light $(C)$ \cite{castro}. 

Many peculiarities in graphene's electronic properties arise from the Dirac nature of the quasi-
particles. One of the examples is the \emph{anomalous quantum Hall effect}. Placed in a magnetic field 
graphene exhibits integer quantum Hall effect, where the Hall states reside at filling factors $\nu=\pm (4 n+2)$ 
with $n=0,1,2,\cdots$ at low magnetic fields ($\sim 10$ T ) \cite{novoselov}. This phenomena can be understood 
from the relativistic nature of the non-interacting quasi-particles. Upon exposing the system to stronger magnetic 
fields ($> 20$ T), additional Hall states appear at filling factors $\nu=0,\pm 1,\pm 4$ \cite{Zhang}. However, the 
plateaus in the Hall conductivity at other odd integer fillings ( for example, $\nu=\pm 3$ or $\pm 5$ etc. ) are absent 
even at the highest laboratory magnetic field ($\sim 45$ T ) \cite{Pkim}. These observations confirm that the four 
fold degeneracy of the zeroth Landau level (LL) is completely lifted. Furthermore, the sub-linear variation of the 
activation gap with magnetic field of the $\nu=1$ Hall state points its origin toward electron-electron interactions. 
Whereas the linear dependence of the activation gap with total magnetic field for $\nu=\pm 4$ Hall states 
suggests that, possibly a finite Zeeman coupling removes only the spin degeneracy from the higher LLs. Within the framework 
of the extended Hubbard model with onsite and nearest-neighbor repulsion, one can show that the valley degeneracy 
only from the zeroth LL can be lifted, due to a spontaneous breaking of the chiral symmetry. In presence of a magnetic field such metal-insulator transition can take place even at sufficiently weak interaction. However, the higher LLs gain a finite shift in energy only. Therefore, one can explain the absence of Hall states at odd integer fillings ( e.g., $\nu=\pm 3, \pm 5$ ) from the protected valley degeneracy of the higher LLs \cite{Miransky,herbut2}.
\\

In addition to the two-dimensional crystal structure, ripples are also present on the graphene sheet and introduce a 
fictitious gauge potential in 
the long wavelength limit \cite{ripples}. Such corrugation of the graphene sheet is partially due to the strain induced by the $SiO_2$ substrate \cite{ishigami}.    
In a normal graphene flake these defects are randomly distributed and therefore 
give a net zero flux of the fictitious field. Perhaps, bulging or stretching of the graphene sheet can give 
rise to a finite pseudo flux \cite{Geimkatguinea}. Recently, its been argued that specific distortions of a flake can give rise to a uniform 
pseudo-magnetic field up to $10$ T \cite{guinea}. Nevertheless, a pseudo-magnetic field of strength $\sim 350$ T, is 
produced by depositing a graphene layer on platinum substrate, followed by cooling of the system \cite{Levy}. In 
contrary to the real magnetic field, the pseudo vector potential couples with opposite sign at two Dirac points and 
hence preserves the time reversal symmetry (TRS). Therefore, in the presence of uniform real and pseudo field, one 
expects the LLs to loose the valley degeneracy. This may also allow the formation of plateaus in Hall
conductivity at odd integer values of $e^2/h$ which were previously absent. \\

Our discussion is concerned with quantization of the Hall conductivity in graphene in the presence of both real and pseudo
magnetic fields. 
The rest of the discussion is organized as follows. In Sec. II, we compute the spectrum of the free Dirac Hamiltonian 
in the presence of both fields. The short ranged electron-electron interaction is discussed in 
Sec. III. Section IV is devoted to the scaling behavior of interaction induced gap. We present the concluding remarks and discuss 
some related issues in Sec. V.   
\\

\section{Free electron spectrum}

The tight binding Hamiltonian for spin-1/2 electrons on the honeycomb 
lattice in the presence of only nearest-neighbor hopping is defined as 
\begin{equation}
H_t=-t \sum_{\vec{A},i,\sigma=\pm} u_{\sigma}^{\dagger}(\vec{A}) 
v_{\sigma}(\vec{A}+\vec{b_i})+ H.c..
\end{equation} 
Here $u^{\dagger}(\vec{A})$ denotes the fermionic creation operator 
on the sites of one of the triangular sublattices generated by the 
linear combination of the basis vectors $\vec{a}_1=(\sqrt{3},-1)a$ and 
$\vec{a}_2=(0,1)a$. $a$ is the lattice spacing ( $\sim 2.5$ $\mathring{A}$ ) 
and $t$ is the nearest-neighbor hopping amplitude $\sim 2.5$ eV  \cite{gloor}. 
Analogously, $v(\vec{A}+\vec{b_i})$ is the electron
annihilation operator on the other sublattice, then located at $\vec{B}
=\vec{A}+\vec{b}$, with the vector $\vec{b}$ being either $\vec{b}_1=
(1/{\sqrt{3}},1)a/2,\vec{b}_2=(1/{\sqrt{3}},-1)a/2$ or $\vec{b}_3=
(-1/{\sqrt{3}},0)a$. Within the
framework of the tight binding model, energy spectrum $E=\pm t |\sum_{i} 
\exp{(i \vec{k} \cdot \vec{b_i})}|$ is doubly degenerate and becomes linear
and isotropic in the vicinity of the corners of the first Brillouin zone. 
Among six such points, only two are inequivalent and suitably chosen to be 
at $\pm \vec{K}$, with $\vec{K}=(1,1/\sqrt{3}) 2\pi/a \sqrt{3}$  \cite{semenoff}.
Hence, retaining the Fourier modes only near these two points, one can write 
down the effective Hamiltonian corresponding to the tight binding model in the 
low energy limit as,
\begin{equation}
H_0=\int d\vec{x} \sum_{\sigma=\pm} \Psi_{\sigma}^{\dagger}(\vec{x}) i\gamma_0
\gamma_i D_i \Psi_{\sigma}(\vec{x}),
\end{equation} 
with
\begin{eqnarray}
\Psi_{\sigma}^\dagger(\vec{x},\tau)= &&{\int^\Lambda} {\frac{d\vec{q}}{(2\pi
a)^2}} e^{- i{\vec{q}\cdot{\vec{x}}}} [u_{\sigma}^\dagger(\vec{K}+\vec{q}),
v_{\sigma}^\dagger(\vec{K}+\vec{q}), \nonumber\\
&& u_{\sigma}^\dagger(-\vec{K}+\vec{q}),
v_{\sigma}^\dagger(-\vec{K}+\vec{q})],
\end{eqnarray}
after conveniently rotating the frame of reference to $q_x={\vec{q} \cdot \vec{K}}/K$ 
and $q_y=(\vec{K} \times \vec{q})\times \vec{K}/K^2$. We adopt natural units, 
$\hbar=e=v_F=1$, where $v_F=ta \sqrt{3}/2$ is the Fermi velocity.
The cut-off in the momentum integral ( $\Lambda \sim 1/a$ ), represents the interval of 
the energy over which the tight binding density of states is
approximately linear. The Einstein summation convention is assumed, but only over 
repeated space-time indices. The mutually anti-commuting four component Hermitian gamma
matrices belong to the ``graphene representation", $\gamma_0=I_2 \otimes \sigma_3, 
\gamma_1=\sigma_3 \otimes \sigma_2$ and $\gamma_2=I_2 \otimes \sigma_1$. We define 
the remaining two gamma matrices as $\gamma_3=\sigma_1 \otimes \sigma_2$ and $\gamma_5=
\sigma_2\otimes\sigma_2$. The low energy Hamiltonian $H_0$ preserves the emergent 
``chiral" $U_c(4)$ symmetry generated by $\{\tau_0,\vec{\tau}\}\otimes\{I_4,\gamma_3,\gamma_5\,
\gamma_{35}\}$ where, $\gamma_{35}=i \gamma_3 \gamma_5=\sigma_3\otimes I_2$ \cite{herbut1,bitan1}. 
The two component Pauli matrices $\{\tau_0,\vec{\tau}\}$, operate on spin indices. 
One can study the response of magnetic field by defining $D_i=-i\partial_i-A_i$, with 
magnetic field $B=\epsilon_{3ij}\partial_i A_j$ set to be perpendicular to the graphene plane. \\  

In addition to the Dirac quasi-particles, a crucial ingredient of the system is ripples. 
The effect of the ripples can be captured in terms of fictitious gauge fields in the 
long wavelength limit. In graphene flakes, ripples are randomly distributed and 
therefore give a net zero flux of pseudo-magnetic field. Whereas, a bulging of the graphene sheet 
may introduce a finite flux of the pseudo-magnetic field. The Dirac Hamiltonian in 
presence of both the real and pseudo-magnetic fields reads as
\begin{equation}
H_0[A,a]=\tau_0 \otimes i \gamma_0\gamma_i (p_i-A_i-a_i^{35}\gamma_{35}),
\end{equation}  
where, $a_i^{35}$ is the member of a general non-Abelian $SU(2)$ gauge field,
\begin{equation}
a_i=a_i^{3}\gamma_3+a_i^{5}\gamma_5+a_i^{35} \gamma_{35}.
\end{equation}
A smooth enough deformation in the graphene sheet, do not mix two inequivalent valleys at $\vec{K}$ and $-\vec{K}$.
Assuming that the bump in the graphene flake varies slowly on the lattice scale we kept only one component of a 
general $SU(2)$ gauge potential \cite{translation}. One way to introduce both gauge potentials in experiment is 
the following. First deposit graphene over a metallic substrate, e.g., platinum \cite{Levy}, at a relatively high   
temperature and then cool the system. Due to a mismatch in compressibility of the substrate and graphene, the former 
one produces strain on the graphene sheet. This way one might expose the system first to a finite pseudo-magnetic field. 
Once the fictitious field is introduced, the system can then be placed in a real magnetic field.     
\\

It is informative to cast the Hamiltonian $H_0[A,a]$ in the following block-diagonal form, 
$H_0[A,a]=H_+[A,a]\oplus H_-[A,a]$, where
\begin{equation}
H_{\pm}=\pm I_2 \otimes \sigma_1 (-i\partial_1-A_1 \mp a_1^{35})-I_2 \otimes \sigma_2 
(-i\partial_2-A_2\mp a_2^{35}).
\label{BDiag}
\end{equation}
$H_{+}$ and $H_-$ represent the Hamiltonian near $\vec{K}$ and $- \vec{K}$ point, respectively.
From Eq.\ [\ref{BDiag}] one can register that the net magnetic field near one Dirac point 
($\vec{K}$) is enhanced to the value $B_{total}^{+}=B+b$, with $b=\epsilon_{3ij}\partial_i
a_j^{35}$. Near the other Dirac point at $- \vec{K}$ the net magnetic field is 
attenuated to $B_{total}^{-}=B-b$. Unless mentioned otherwise we assume $B > b$ for the 
rest of our discussion. Both the fields are assumed to be uniform as well. One can see that 
both $H_{\pm}$ are unitarily equivalent to a generic Dirac Hamiltonian $H_D[A,a]$ in two dimensions 
in the presence of gauge fields, where
\begin{equation}
H_D[A,a]=i \gamma_0 \gamma_i (p_i-A_i-a^{35}_i).
\end{equation} 
Specifically, $H_+=U^\dagger_1H_D[A,a]U_1$, with $U_1=I_2\oplus i\sigma_2$ and $H_-=U^\dagger_2H_D[A,-a]U_2$, 
with $U_2=i\sigma_2 \oplus I_2$ \cite{herbutAF}. \\

In the absence of the pseudo-magnetic field $(b=0)$ the Hamiltonian $H_0[A,0]$, exhibits a series of LLs 
at well separated energies $\pm \sqrt{2nB}$, $n=0,1,2,\cdots $, with degeneracies 
of $B/\pi$ per unit area. At half-filling, all the negative energy LLs are completely filled, 
whereas the LLs at positive energies are totally empty, with only half of the zero energy states
occupied. As a consequence, the Hall conductivity shows plateaus at integer fillings $\nu=\pm(4n+2)$,
$ n=0,1,2,\cdots $ \cite{Gusynin}. Measurement of the Hall conductance at relatively low magnetic fields
($B \sim 10$ T) confirmed such quantization \cite{novoselov,Zhang}. The additional four-fold degeneracy of the 
LLs arises from the spin and the valley degrees of freedom. The same story follows when one switches 
off the real magnetic field, and only a pseudo-magnetic field penetrates the system, with the 
difference that now the LLs appear at energies $\pm \sqrt{2nb}$. A recent experiment confirmed such quantization
in the presence of a pseudo-magnetic field of strength $\sim 350$ T \cite{Levy}.
However, the presence of finite real and pseudo-magnetic field removes the valley
degeneracy from each LL. 
Consequently, each of the LLs at energy $\sqrt{2nB}$, upon imposing a finite pseudo-magnetic field, 
gives rise to  two new LLs at energies $\sqrt{2 n(B\pm b)}$, with degeneracies $D_{\pm}=(B\pm b)/2\pi$ per 
unit area, respectively, but only for $n \neq 0$. It is worth noticing that the Dirac points are decoupled 
from each other even in presence of a finite pseudo field. Hence, there are two sets of zero energy states, 
one is localized near the $\vec{K}$ point, and the other one near $-\vec{K}$ (or $\vec{K'}$) with degeneracies 
$\Omega(B\pm b)/2\pi$,
respectively. Here $\Omega$ being the area of the sample. Thus, the zeroth LL does not split even when the system 
experiences finite real and pseudo fields \cite{explan2,jackiw}. However, in presence of uniform 
real and pseudo-magnetic field the higher LLs split, thereby pushing the states near $K$ point up in energy, and 
those in the vicinity of $K'$ point down in energy. Hence, with some particular strength of $B$ and 
$b$, states from two successive LLs, localized near different Dirac points can be degenerate. 
This situation can easily be bypassed by tuning the ratio $B/b$ close to an \textit{even integer}. \\

Upon including a finite Zeeman coupling, the spin degeneracy can be lifted from all the LLs including the 
zeroth one. The Zeeman splitting scales as $\Delta_Z$ (in \textrm{Kelvin}) $\sim B$, where $B$ is measured in 
\textrm{Tesla}, thus much smaller than the LL energies. For our purposes we tune the magnetic fields so 
that, $5<B/b<10$. Within this range of parameters, the energy difference between the Hall states with same LL index 
$(n)$ but localized near two different valleys is $\sim 200-400$ K. Energies corresponding to different LLs for 
one particular set of realistic values of the magnetic fields are shown in Table I. Without considering 
the many body effects arising from the electron-electron interactions, one can expect the Hall 
conductivity to exhibit plateaus at integer values $f=0,\pm 2,\pm 3,\pm 4,\pm 5,\cdots$ of $e^2/h$. 
It is worth noticing that the LLs associated with different Dirac points do not enjoy the luxury of equal degeneracy. 
Therefore, plateaus of Hall conductivity at integer values of $e^2/h$, are not associated with the integer fillings.  
For further illustration of the computation of the Hall conductivity the reader may refer to the Appendix. \\ 

\begin{table}[ht]
\caption{Energies of LLs with $B=32$ T and $b=4$ T} 
\centering
\begin{tabular}{c c c c}
\hline\hline
n \hspace*{0.2cm} & 1 \hspace*{0.2cm} & 2 \hspace*{0.2cm} & 3 \\[0.5ex]
\hline \vspace*{0.1cm}\\
$E_0$ \hspace*{0.2cm} & $2.4\times 10^3$ K \hspace*{0.2cm} & $3.4 \times 10^3$ K \hspace*{0.2cm} & $4.1 \times 10^3$ K \vspace*{0.5cm} \\
$E_+$ \hspace*{0.2cm} & $2.5\times 10^3$ K \hspace*{0.2cm} & $3.5 \times 10^3$ K \hspace*{0.2cm} & $4.4 \times 10^3$ K \vspace*{0.5cm} \\
$E_-$ \hspace*{0.2cm} & $2.2\times 10^3$ K \hspace*{0.2cm}  & $3.2 \times 10^3$ K \hspace*{0.2cm}  & $3.8 \times 10^3$ K \vspace*{0.3cm}\\
\hline
\end{tabular} \vspace*{0.1cm}\\
{\footnotesize Energies of the LLs (measured from the charge neutral point) for a particular choice of 
the magnetic fields. Here $n$ stands for the LL index in absence of pseudo field. $E_0$ stands for the 
energy of the LL without any pseudo field. $E_{+ (-)}$ denotes the energy of the LL lives in the vicinity of the 
$+\vec{K} (-\vec{K})$ Dirac point.}
\label{LLen}
\end{table}

\section{Electron-electron interactions}

Diagonalizing the Dirac Hamiltonian in the presence of uniform real and pseudo-magnetic fields we found that
the valley degeneracy is lifted only from higher LLs. In the absence of interactions the zeroth LL still enjoys the 
valley degeneracy. In this section we will consider the effect of the short-range electron-electron 
interactions in the spectrum. The interacting Hamiltonian in presence of only on-site (U) and nearest-neighbor 
(V) repulsion is defined as,
\begin{equation}
H_U=\frac{U}{2} \sum_{\vec{X},\sigma} n_{\sigma}(\vec{X}) n_{-\sigma}(\vec{X}) +\frac{V}{2}\sum_{\vec{A},i,
\sigma,\sigma'} n_{\sigma} (\vec{A})n_{\sigma'}(\vec{A}+\vec{b_i}).
\label{hubbard}
\end{equation}  
For graphene, $U\approx 5-12$ eV and $U/V \approx 2-3$ \cite{hoffmann}. We consider the system to be at
filling one half. In the absence of magnetic fields, a 
sufficiently large on-site interaction can take the system into an insulating ground state with the magnetization
altering its sign at each site from its neighbor. A staggered pattern in average electron density
can be realized at large enough nearest-neighbor Coulomb repulsion. The transitions out of the semi-metal to the 
Mott insulators are believed to be continuous and belong to the Gross-Neveu universality class \cite{herbut1,bitan1}. 
Therefore, we assume a uniform background of either the staggered density [$C=\langle n(\vec{A})-n(\vec{A}+\vec{b}
)\rangle$] or staggered magnetization [$N=\langle m(\vec{A})-m(\vec{A}+\vec{b})\rangle$]. Here $n(\vec{A})=u^\dagger_\sigma
(\vec{A}) u_\sigma(\vec{A})$ is the average electron density and $m(\vec{A})=u^\dagger_\sigma(\vec{A})\tau_3 u_\sigma(\vec{A})$ corresponds to average magnetization on sublattice A. Similar quantities are analogously defined on B sublattice in terms of fermionic operators $v_\sigma(\vec{B})$ and $v^{\dagger}_\sigma(\vec{B})$. After the usual Hartree-Fock 
decomposition, one can write down an effective single-particle Hamiltonian in that background as 
\begin{equation}
H_{HF}=\tau_0 \otimes H_0[A,a] + m a\otimes \gamma_0,
\end{equation} 
where $a=\tau_0$ corresponds to $m=C$ [ charge density wave (CDW)] and $a=\tau_3$ to $m=N$ [ anti-ferromagnet 
(AF)]. Here we omitted the Zeeman term for convenience. The effect of Zeeman splitting will be discussed later. \\

The spectrum of the Hamiltonian $H_{HF}$ is as follow: For $n\neq 0$ the eigenvalues are at $\pm \sqrt
{2n(B\pm b)+m^2}$, for each spin projection, with degeneracies per unit area $D_{\pm}=(B\pm b)/2\pi$, respectively. 
Besides these, for $n=0$, the eigenvalues of $H_{HF}$ for each spin components are $\pm |m|$ with degeneracies $D_{\pm}^0=(B\pm b)/2\pi$ per unit area, respectively. 
Therefore filling up only the states at negative energy  while leaving those at positive energy empty one immediately 
develops a gap even at small enough interaction. Generating a gap at infinitesimal interactions in the presence of a magnetic field is termed as ``magnetic catalysis" \cite{gusynin2}. This has been proposed as a mechanism behind formation of the Hall states in graphene at filling factor $\nu=0$ and $1$ in presence of an ordinary magnetic field \cite{Miransky,herbut2}. In the zeroth LL, states associated
with $\vec{K}$ and $-\vec{K}$ are localized on complimentary sublattices A and B respectively. Hence,
for each spin projection, there are $(B + b)/2\pi$ states per unit area on the A sublattice, whereas the other sublattice hosts only $(B-b)/2\pi$ states. In the absence of pseudo-field, states on
each of the sublattices enjoy equal degeneracy. Therefore, even infinitesimally strong interactions can develop a gap, at filling one-half by spontaneously breaking the chiral symmetry of the Dirac quasi-particles.   
However, the nature of the insulating ground state depends on the strength of the interactions at the lattice scale. For example, if $U \gg V$, AF ordering lowers the the energy of the ground state. A CDW order can develop within the zeroth LL if on the other hand, the nearest-neighbor component of the finite range Coulomb repulsion is the dominant one. However, due to imbalance in the number of states living on two sublattices in the zeroth LL,
only the \textit{ferrimagnetic} order can be developed  while keeping the system at charge neutrality. The electron-electron interactions remove either the valley degeneracy or the sublattice degeneracy from the zeroth LL. These two are equivalent however only within the zeroth LL.  Alternatively, a large enough Zeeman coupling will orient the magnetization at all the sites in the direction of the magnetic field and hence lower the energy of the filled Dirac-Fermi sea. Such a ground state with finite magnetization can also be stabilized by increasing the component of the magnetic field parallel to the graphene plane. The exact nature of the ground state at $\nu=0$ in graphene in presence of only real field is still a subject of debate \cite{Fuchs,herbutSO3,Fisher,McDonald}. The activation gap of longitudinal resistivity for  $\nu=1$ was found to be independent of the component of the magnetic field parallel to the graphene plane, and varies sublinearly  with the its perpendicular component. This observation strongly suggests that it is originating from the electron-electron interactions \cite{Zhang2}.
 \\

Therefore upon including the Zeeman splitting and the finite ranged interactions the four fold degeneracy from all the LLs is completely removed. Within the zeroth LL, a gap develops in by lifting the valley or sublattice degeneracy. Under this circumstance, one can expect the Hall conductivity to exhibit quantized plateaus at 
$\sigma_{xy}=f e^2/h$, with $f=0,\pm1,\pm2,\pm3,\cdots$ \cite{append}. \\

\section{Scaling of interaction induced gap}

Let us now focus on the scaling behavior of the interaction induced gap. Validity of our theory relies on the assumption that the magnetic fields are low in comparison to the characteristics lattice magnetic field, or equivalently the magnetic length is much larger than the lattice spacing. This condition is easily satisfied even at the highest laboratory magnetic field $B \sim 45$T and $0<b/B<0.2$. In a current experimental situation with $b\sim 350$ T \cite{Levy}, the magnetic length ($\approx 35 \mathring{A}$) is more than an order magnitude larger than the lattice spacing ($a\sim 2.5 \mathring{A}$). Hereafter we assume that the spin degeneracy is completely lifted by the Zeeman splitting and the chemical potential lies close to the Zeeman shifted `Dirac' point, thus $f=1$. After integrating out the fast Fourier modes within the momentum shell $1/a<k<1/l_B$, one can write down the effective low energy Lagrangian corresponding to $H_U$ in Eq.\ [\ref{hubbard}] as
\begin{eqnarray}
L&=&i\sum_{\sigma}\overline{\Psi}_{\sigma} \gamma_{\mu}D_{\mu} \Psi_{\sigma}-g_1\left( \sum_\sigma \overline{\Psi}_\sigma \Psi_\sigma \right)^2 \nonumber \\
&-& g_2 \left( \sum_\sigma \sigma\overline{\Psi}_\sigma \Psi_\sigma \right)^2,
\end{eqnarray}
where $\overline{\Psi}_{\sigma}=\Psi_{\sigma}^\dagger  \gamma_0$, $D_0=-i\partial_{\tau}$, and $\tau$ is the imaginary time. 
$\mu=0,1,2$ runs over the space-time indices and summation over repeated indices is assumed. Here, $l_B \sim B^{-1}$ is the magnetic length and we kept only the least irrelevant couplings $g_1=(3V-U)a^2/8$ and $g_2=U a^2/8$. Assuming a uniform background of either staggered electron density or \emph{ferrimagnetic} order, the ground state energy for the $N$ number of four component fermions in the magnetic fields is given by
\begin{widetext} 
\begin{equation}
\frac{E(m)-E(0)}{N}=\frac{m^2}{4g} + \sum_{\sigma=\pm}\frac{B+\sigma b}{4\pi^{3/2}} \int^{\infty}_0 \frac{ds}{s^{3/2}} 
(e^{-sm^2}-1) [\frac{1+\sigma}{2}+  K(s\Lambda^2) (\coth(s (B+\sigma b))-1) ],
\label{freeEn}
\end{equation}
\end{widetext}
as $N \rightarrow \infty$. Setting $b=0$, one recovers the ground state energy in the presence of only real magnetic 
field \cite{herbut2}. Here $K(x)$ is the cut-off function introduced to sum over the $n\neq 0$ LLs. This function satisfies $K(x \rightarrow \infty)=1$ and $K(x \rightarrow 0 )=0$, but otherwise arbitrary. $E(m)$ can be understood as the Hartree-Fock variational ground state energy of the 
electrons either in a CDW or ferrimagnetic background. Due to the different degeneracies of the states localized near $\vec{K}$ 
and $\vec{K'}$, the energy of the ground state is maximally lowered by pushing down all the states on $A$ sublattice 
below the chemical potential while leaving those on $B$ sublattice empty. In the absence of pseudo flux, 
$A$ and $B$ sublattice hosts equal number of states. System then spontaneously chooses the ground state by breaking the
Ising like symmetry of either sublattice or the valley degrees of freedom. The first term in the parentheses in 
Eq.\ [\ref{freeEn}] counts the contribution from the zeroth LL whereas, the second one includes the contributions from 
higher LLs. Minimizing $E(m)$, one can cast the gap equation in the following form
\begin{equation}
\frac{X}{2}= f(X,q)+  \frac{\delta}{m},
\end{equation} \\
where $X=(B+b)/m^2$, $\delta=(g\Lambda)^{-1}-(g_c \Lambda)^{-1}$ measures the deviation from the critical interaction 
($g_c$) and $q=(B+b)/(B-b)$, yielding $B/b=(q+1)/(q-1)$. The non-universal value of the critical interaction is 
determined by
\begin{equation}
\frac{1}{g_c}=\frac{\Lambda}{\sqrt{\pi}} \int_{0}^{\infty} ds \frac{K(t)}{s^{3/2}},
\end{equation} 
and the function $f(X,q)$ is defined as
\begin{widetext}
\begin{equation}
f(X,q)=\frac{1}{\sqrt{\pi}} \int_{0}^{\infty} \frac{ds}{s^{3/2}} K\left(\frac{s X B_0}{B+b}\right) 
\left(1- \frac{X s e^{-s}}{e^{2 X s}-1}-\frac{1}{q} \cdot \frac{X s e^{-s}}{e^{2 X s/q}-1} \right).
\end{equation}
\end{widetext}
In the limit of the low magnetic fields $B,b \ll 
B_0$, where $B_0\sim a^{-2}$ is the magnetic field corresponding to the lattice scale, one can substitute $K(x)$ by 
\textit{unity}. Upon setting $q=1$, one gets the gap equation in presence of the ordinary magnetic field, 
obtained previously \cite{bitan2}. However, for $1 \leq q \leq 1.5 $, or equivalently $ 0\geq b/B \geq 0.2$, the solution of the gap equation lies in the range $17.913 \geq X=X_0 \geq 14.0$, when $\delta=0$. Realizing that the solution of the gap equation for $\delta >0$ or $g < g_c$ will exist at $X> X_0$, one can expand $f(X,q)$ for large $X$, at least when $q$ is not far from \emph{unity}. Keeping the terms up to the second order one can write
\begin{equation}
f(X,q)=u \left(1+\frac{1}{\sqrt{q}} \right) \sqrt{X}+\frac{v}{\sqrt{X}} \left(1+\sqrt{q} \right) +O(X^{-3/2}),
\end{equation} 
with $u=1.03258$ and $v=0.461808$. Hence one can cast the self consistent equation of the gap in terms of an algebraic equation
\begin{widetext}
\begin{equation}
u \sqrt{\frac{2q}{q+1}}\left(1+\frac{1}{\sqrt{q}}\right) m 
+v \frac{ \sqrt{1+q}}{\sqrt{2q}} (1+\sqrt{q}) \frac{m^3}{B}
+\frac{ \delta}{\sqrt{ B}} m -\frac{q}{1+q}\sqrt{B}=0.
\end{equation}
\end{widetext}
At criticality, i.e., $\delta=0$, the interaction-induced mass varies as $m=\sqrt{2B}/C$, where $C$ now depends on $q$,
otherwise it is a universal number. Particularly for $q=1$, i.e., $b=0$, this ratio was found to be $C=5.985+O(1/N)$ \cite{bitan2}. Recently the same universal number has been computed numerically on a finite honeycomb lattice, where we found this number to be extremely close to the one computed in continuum limit \cite{inhomog}. From the solution of the gap equation at $\delta=0$, one finds that the size of the gap at $f=1$ increases upon introducing a finite pseudo flux. This
issue can be resolved by considering the zeroth LL only. In the presence of real and pseudo fields, Hall conductivity develops a plateau at $f=1$ by filling $\Omega (B+b)/2\pi$ states from the charge neutrality, where $\Omega$ is the area of the sample. Hence the zeroth LL contribution to the ground state energy is proportional to $\Omega m (B+b)/2\pi$, where $m$ is the activation gap for $f=1$.
Therefore, one finds that the zeroth LL contribution is enhanced in the presence of pseudo flux. In the gap equation this contribution is dominant whereas, the higher LL contributions are $O(m^2/B_{\pm})$ and hence sub-dominant. After some tedious, otherwise 
straight-forward calculations it can be shown that the higher LL contribution also increases the size of the gap at $f=1$.
We computed the gap as a function of $B$ for $q=1$ and $1.285$ for various $\delta$ and a family of such curves is shown in Fig. 1.  The first case, $q=1$ corresponds to $b=0$ and $q=1.285$ implies $B/b\approx 8$. 
\\

\begin{figure}[t]
{\centering\resizebox*{80mm}{!}{\includegraphics{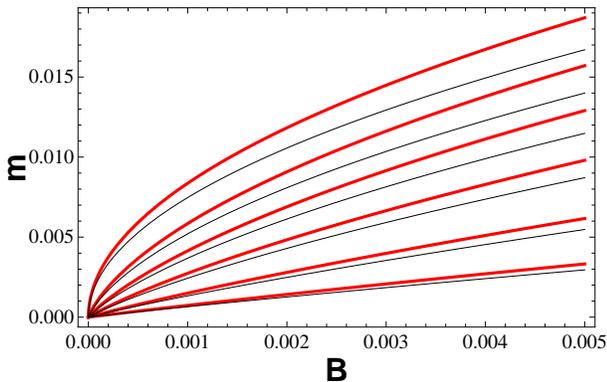}}
\par} \caption[] {(Color online) Gap ($m$) at $f=1$ as function of $B$, with $q=1.285$ [red (thick)], i.e., $B/b=8$ and $q=1$ (black), i.e., $b=0$. Here $m$ and $B$ are measured in units of $v_F \Lambda$ and $B_0=\Lambda^2$, respectively. The top curves correspond to the critical point $\delta=0$ and the remaining ones to $\delta=0.03,0.07,0.14,0.31,0.7$ (top to bottom), where $\delta=(g_c-g)/g_cg \Lambda$.  }
\end{figure} 

The gap at $f=1$ is generated by breaking either sublattice or the valley degeneracy within the zeroth LL.
The semimetal-insulator transition belongs to the Gross-Neveu universality class. Therefore, near transition 
the dynamical critical exponent $\emph{z}$ is equal to \textit{unity} and the correlation length exponent $\nu=1+O(1/N)$ \cite{bitan1}. The 
$1/N$ corrections are found to be small in comparison to the leading order contribution. This is the reason behind 
their omission here \cite{wetterich}. One can also confirm the critical exponents from the linear variation of the 
gap with magnetic field $B$, for small interactions $(\delta \ll 0)$ \cite{ref}. As one enters the regime of stronger 
interaction, a sub-linear dependence of the gap on magnetic field emerges. At criticality $\delta=0$, the gap shows a 
perfect square-root dependence on magnetic field \cite{bitan2}. For $q=1.285$, one finds the universal ratio $C$ to 
be $5.55878+O(1/N)$. To the leading order, the scaling function does not depend on the exact nature of the short-ranged interactions. Any order parameter that lifts the valley degeneracy from the zeroth LL 
will lead to identical scaling behavior. However, $1/N$ corrections may depend on the exact nature of the order parameters 
\cite{herbut1,oskar}.  
\\

Pseudo-magnetic fields may arise from specific deformation of the graphene sheet, thus field profile is typically \emph{inhomogeneous}. Hence it natural to study the behavior of the collection of the interacting fermions on honeycomb lattice when the fields are inhomogeneous. A 
recent numerical study \cite{inhomog} in the presence of a nonuniform, but real magnetic field, established that the catalysis mechanism survives even when the condition of uniformity of the field is relaxed. System finds itself in an ordered phase at sufficiently weak interaction by developing a local expectation value of the order parameter, which more or less is found to follow the profile of the magnetic field. System finds itself in an ordered phase by lifting the chiral symmetry. A similar behavior of the \emph{local} order parameter on the \emph{local} strength of the field is been observed, when only a inhomogeneous pseudo-magnetic field penetrates through the system
\cite{pseudo}. The order parameter in presence of fictitious field breaks the TRS \cite{herbut3}. Hence, in presence of inhomogeneous fields, we will expect the order parameter to develop a space modulated expectation value. However, we leave this issue for further investigation. \\

\section{Summary and discussion}

To summarize, we consider Dirac fermions in two spatial dimensions in the presence of real and
pseudo-magnetic fields. For simplicity, we assumed both of them to be uniform and the real one is directed 
perpendicular to the graphene plane. Diagonalizing the free Hamiltonian under that circumstance we found that all 
the LLs at finite energy lack valley degeneracy, whereas the one at zero energy enjoys it. Taking into
account the onsite and nearest-neighbor interaction we have shown that, the zero energy level splits by 
developing either a CDW or ferrimagnetic order when the real magnetic field is stronger than the fictitious 
one. Upon including the Zeeman splitting, the spin degeneracy from all the LLs is lifted and
one expects to see plateaus in the Hall conductivity at all the integer factors of $e^2/h$. Previously 
the plateaus at odd integer fillings $f=\pm 3,\pm 5,\cdots$ could not be observed due to the valley 
degeneracy of the higher LLs. Here we proposed that a finite pseudo flux can, in principle, release the 
LLs from the valley degeneracy. However, due to distinct degeneracies of the LLs localized in the vicinity of two Dirac points,
the plateaus in Hall conductivity at integer values of $e^2/h$, would be observed at non-integer fillings. A scaling behavior of the interaction induced gap at $f=1$ is presented, 
assuming either onsite Hubbard or nearest-neighbor repulsion is the dominant finite range interaction. Within the mean field approximation, the size of 
the gap is found to be enhanced in the presence of a finite pseudo flux. One way to test the enhancement of the gap is as follows. As mentioned in Ref. \cite{Levy}, the bulging of the graphene sheet does not take place everywhere on the sample. Only some portion of the flake develops a bump and experiences a finite pseudo field. Therefore placing the system in a magnetic field one can measure the order parameter in two different domains of the flake, where the flake is bulged and the other where it is reasonably flat. This way one can determine the effect of the finite pseudo flux on the interaction driven gap.
\\

In a finite system, pseudo field can be achieved by means of an inhomogeneous strain. Particular deformations  
that might give rise to such field was proposed in Ref. \cite{Geimkatguinea} and in Ref. \cite{guinea}. One feature of such deformation is that it leaves the lattice only with a $C_3$ symmetry, which has also been confirmed experimentally \cite{Levy}. As long as the extension of the bump in graphene flake and the magnetic length is larger than the lattice spacing, continuum description is valid. In recent experiment  
\cite{Levy}, $l_b\approx 35 \mathring{A}$ at $b=350$ T and the extension of the bump is $\sim 40-100 \mathring{A}$. 
Hence, both the length scales are much larger than the lattice spacing in graphene ($a \approx 2.5 \mathring{A}$).
However, for true thermodynamic limit, one requires the extension of the bump to be much larger than the magnetic length.
In present experimental situation these two length scales are comparable. However, with a smoother bump with broader extension, one can get to the regime where the constrain is easily satisfied.   
\\

Finally let us turn to the situation when $b> B$. Recently a \textit{uniform} pseudo field of strength $350$ T was 
produced by bulges in the graphene layer deposited on Platinum substrate \cite{Levy}. Hence this limit seems quite 
achievable. Particularly for $B=0$ the zero energy states near two Dirac points live on the same sublattice. 
Hence, an infinitesimal next-nearest-neighbor Coulomb repulsion may open up a gap by spontaneously breaking time 
reversal symmetry (TRS) \cite{herbut3}. The order parameter in the insulating phase is given by $\langle {\Psi^
\dagger_\sigma ( I_2 \otimes i \gamma_1\gamma_2 ) \Psi_\sigma}\rangle$. The TRS is represented by an anti-unitary 
operator $I_t=U_t K$, where $U_t$ is unitary and $K$ is complex conjugation operator. In the graphene representation 
$U_t$ is given by $i \gamma_1 \gamma_5=\sigma_1 \otimes I_2$ \cite{bitan1}. The correlated ground state under this 
circumstance corresponds to an intra sublattice Haldane circulating current, propagating in opposite direction on 
the two sublattices \cite{Haldane}. This state is named the \emph{quantum anomalous Hall insulator} (QAH).
The exact nature of the correlated ground state in presence of a pseudo-magnetic field is yet to be determined and 
we leave this issue for future investigation.
However, in the absence of any gauge potentials ($B=0, b=0$) fluctuations preempt the appearance of QAH and stabilizes 
the \emph{spin Hall insulator} (QSH) ground state. The correlated 
ground state in the QSH phase, breaks the TRS for each spin component and the order parameter reads as $\langle {\Psi^
\dagger_\sigma (\vec{\tau} \otimes i \gamma_1 \gamma_2) \Psi_\sigma}\rangle$ \cite{raghu}. 
  \\

Here we focused only on the finite-ranged components of the Coulomb interaction. On the other hand, 
its long range $1/r$ tail is unscreened due to the vanishing density of states at the charge neutral 
point. If one turns off the pseudo-magnetic field the long range Coulomb interaction will lead to 
a similar splitting of the LLs. However, it introduces an energy scale $e^2/\epsilon l_B$, 
which is on the same order of the LL energy with commonly assumed dielectric constant $\epsilon \approx 
5$ for $SiO_2$ substrate. This immediately contradicts the experimentally observed gap of $\sim 100$ K at $f=1$, an order of magnitude smaller than the LL energy \cite{Zhang2}. This issue can be resolved 
by assuming an order of magnitude larger value of the dielectric constant, which attributes the accumulation of water 
layer between the graphene flake and  $SiO_2$ substrate \cite{schedin}. In recent work \cite{Haldane2}, 
it was shown that even though the long range Coulomb interaction liberates all the LLs from the valley degeneracy,
the critical strength of the disorder above which the $f=3$ plateau is destroyed is much smaller than that for 
$f=1$. This scenario, may explain the absence of odd integer Hall plateaus in graphene \cite
{Zhang,Zhang2}. In contrast, the presence of a finite flux of pseudo field leads to an activation gap $\sim 200
-400$ K for $f=\pm 3,\pm 5$ Hall states (Table I). Consequently, the Hall states at those fillings should be much more robust 
against disorder. Thus, one may expect to see the quantization of Hall conductivity at odd integer factors of $e^2/h$ in a 
cleaner system. Assuming a sufficiently large dielectric constant, one can restrict oneself 
to the short-ranged components of the Coulomb interaction. Effect of weak long-ranged
Coulomb interaction on the interaction mediated gap is discussed in Ref. \cite{bitan2}.   
\\

\section{Acknowledgments}

Author is grateful to Igor Herbut and Malcolm Kennett for discussions and critical reading of the manuscript.
Author is also grateful to Somshubra Sharangi for useful suggestion about the manuscript.    
This work was supported by NSERC of Canada. \\

\section{Appendix}

Here we determine the quantization of the Hall conductivity that we mentioned at the end of the Sec. II and Sec. III. One notices that upon introducing a finite pseudo-magnetic flux, all the LLs do not enjoy equal degeneracies. LLs localized in the neighborhood of the Dirac point at $\vec{K}$, have a degeneracy $(B+b)/2\pi$ per unit area, whereas those living in the vicinity of the other Dirac point, at $-\vec{K}$, carry $(B-b)/2\pi$ states per unit area. Using the St\v{r}eda formula, one can write the expession of the Hall conductivity as
\begin{equation}
\sigma_{xy}= \left( \frac{\partial N}{\partial B} \right)_{\mu},
\end{equation}
after setting $e=c=1$ \cite{streda}. Here $N$ is the electronic density in the bulk and the derivative with respect to $B$ is taken at fixed chemical potential ($\mu$). The chemical potential is measured from the charge neutral point. 
Moreover, all the LLs have additional two fold spin degeneracy. Upon changing the magnetic field $B$, number of states below the chemical potential
is changed to 
\begin{equation}
\delta N =\Omega f_+ \delta B + \Omega f_-  \delta B.
\end{equation}
$f_+$ and $f_-$ counts the number of LLs below the chemical potential ($\mu$), but above the charge neutral point, with degeneracies $(B\pm b)/2\pi$ per unit area for each spin species, respectively. Therefore the Hall conductivity is
\begin{equation}
\sigma_{xy}=f_+ +f_-=f.
\end{equation} 
Hence, quantization of the Hall conductivity only takes the number of filled LLs below the chemical potential into account. That leads to the the announced result of the Hall conductivity in presence of real and pseudo-magnetic fields.
However, one should notice that due to the different degeneracies of the LLs, quantization of the Hall conductivity at integer values of $e^2/h$ is not associated with integer fillings. The filling factors explicitly depend on the ratio $b/B$.  
 \\


\begin{thebibliography}{99}
\bibitem{Novoselov}
K. S. Novoselov, A. K. Geim, S. V. Morozov, D. Jiang, Y. Zhang, S. V. Dubonos, I. V. Grigorieva 
A. A. Firsov, Science, {\bf 306}, 666 (2004).
\bibitem{castro}
A. H. Castro Neto, F. Guinea, N. M. R. Peres, K. S. Novoselov, and A. K. Geim, Rev. Mod. Phys. {\bf 81}, 109 (2009). 
\bibitem{novoselov}
K. S. Novoselov, A. K. Geim, S. V. Morozov, D. Jiang, M. I.
Katsnelson, I. V. Grigorieva, S. V. Dubonos, and A. A. Firsov,
Nature (London) {\bf 438}, 197 (2005͒),
Y. Zhang, Y.-W. Tan, H. L. Stormer, and P. Kim, Nature (Lon-
don) {\bf 438}, 201 (2005).
\bibitem{Zhang}
Y. Zhang, Z. Jiang, J. P. Small, M. S. Purewal, Y.-W. Tan, M. Faziollahi, J. D. Chudow,
J. A. Jaszczak, H. L. Stromer, and P. Kim, Phys. Rev. Lett. {\bf 96}, 136806 (2006).
\bibitem{Pkim}
See however, C. R. Dean, A. F. Young, I. Meric, C. Lee, L. Wang,
S. Sorgenfrei, K. Watanabe, T. Taniguchi, P. Kim, K. L. Shepard  and J. Hone, Nat. Nanotechnology,
{\bf 5}, 722 (2010).
\bibitem{Miransky}
V. P. Gusynin, V. A. Miransky, S. G. Sharapov and I. A. Shovkovy, Phys. Rev. B {\bf 74}, 195429 (2006).
\bibitem{herbut2}
I. F. Herbut, Phys. Rev. B {\bf 75}, 165411 (2007).
\bibitem{ripples}
See Sec. IV of Ref. \cite{castro}.
\bibitem{ishigami}
M.Ishigami, J. H. Chen, W. G. Cullen, M. S. Fuhrer, and E. D. Williams, Nano Lett. {\bf 7}, 1643 (2007).
\bibitem{Geimkatguinea}
F. Guinea, M. I. Katnelson and A.K. Geim, Nat. Phys. {\bf 6}, 30 (2010).
\bibitem{guinea}
F. Guinea, A. K. Geim, M. I. Katnelson, K. S. Novoselov, Phys. Rev. B {\bf 81}, 035408 (2010). 
\bibitem{Levy}
N. Levy, S. R. Burke, K. l. Meaker, M. Palnasigui, A. Zettl, F. Guinea, A. H. Castro Neto, M. F. Crommie, Science {\bf 329}, 544 (2010).
\bibitem{gloor}
T. A. Gloor and F. Milla, Eur. Phys. J. B {\bf 38}, 9 (2004).
\bibitem{semenoff}
G. W. Semenoff, Phys. Rev. Lett. {\bf 53}, 2449 (1984).
\bibitem{herbut1}
I. F. Herbut, Phys. Rev. Lett. {\bf 97}, 146401 (2006).
\bibitem{bitan1}
I.F. Herbut, V.Juri\v{c}i\'{c}, B. Roy, Phys. Rev. B {\bf 79}, 085116 (2009).
\bibitem{translation}
Two other components of the fictitious gauge potential, proportional to $\gamma_3$ and $\gamma_5$ are
off diagonal in valley index, hence mixes two inequivalent valleys. 
\bibitem{herbutAF}
I. F. Herbut, Phys. Rev. Lett. {\bf 99}, 206404 (2007).
\bibitem{Gusynin}
V. P. Gusynin and S. G. Sharapov, Phys. Rev. Lett. {\bf 95}, 146801 (2005);
N. M. R. Peres, F. Guinea and A. H. Castro Neto, Phys. Rev. B {\bf 72}, 174406 (2005).
\bibitem{explan2}
There is one copy of 4-component Dirac fermions near each of the Dirac points. Hence each of them hosts zero energy modes.
\bibitem{jackiw}
R. Jackiw, Phys. Rev. D {\bf 29}, 2375 (1984).
\bibitem{hoffmann}
A. L. Tchougreeff and R. Hoffmann, J. Phys. Chem. {\bf  96} 8993 (1992).
\bibitem{gusynin2}
V. P. Gusynin, V. A. Miransky, and I. A. Shovkovy, Phys. Rev. Lett. {\bf 73}, 3499 (1994), Phys. Rev. D {\bf 52}, 4718 (1995).
\bibitem{Fuchs}
J. -N. Fuchs and P. Lederer, Phys. Rev. Lett. {\bf 98}, 016803 (2007).
\bibitem{Fisher}
J. Alicea, M. P. A. Fisher, Phys. Rev. B {\bf 74}, 075422 (2006).
\bibitem{herbutSO3}
I.F. Herbut, Phys. Rev. B {\bf 76}, 085432 (2007).
\bibitem{McDonald}
J. Jung, A. H. MacDonald, Phys. Rev. B {\bf 80}, 235417 (2009).
\bibitem{Zhang2}
Z. Jiang, Y. Zhang, H. L. Stromer, P. Kim, Phys. Rev. Lett. {\bf 99}, 106802 (2007).
\bibitem{append}
Consult Appendix.
\bibitem{bitan2}
I.F. Herbut, B. Roy, Phys. Rev. B {\bf 77}, 245438 (2008).
\bibitem{inhomog}
B. Roy and I. Herbut, Phys. Rev. B {\bf 83}, 195422 (2011).
\bibitem{wetterich} 
L. Rosa, P. Vitale, and C. Wetterich, Phys. Rev.
Lett. {\bf 86}, 958 (2001); F. H\"{o}fling, C. Novak, and C.
Wetterich, Phys. Rev. B {\bf 66}, 205111 (2002).
\bibitem{ref}
For details consult Ref. \cite{bitan2}
\bibitem{oskar}
I. F. Herbut,  V.Juri\v{c}i\'{c}, O. Vafek, Phys. Rev. B {\bf 80}, 075432 (2009); B. Roy, e-print arxiv:1106.1419 (to be published).
\bibitem{pseudo}
B. Roy and I. Herbut, unpublished.
\bibitem{herbut3}
I.F. Herbut, Phys. Rev. B {\bf 78}, 205433 (2008).
\bibitem{Haldane}
F. D. M. Haldane, Phys. Rev. Lett. {\bf 61}, 2015 (1988).
\bibitem{raghu}
S. Raghu, Xiao-Liang Qi, C. Honerkamp, Shou-Cheng Zhang, Phys. Rev. Lett. {\bf 100}, 156401 (2008).
\bibitem{schedin}
F. Schedin, A. K. Geim, S. V. Morozov, D. Jiang, E. H. Hill, P. Blake, K. S. Novoselov, Nat. Mater. {\bf 6}, 652 (2007͒).
\bibitem{Haldane2}
L. Sheng, D. N. Sheng, F. D. M. Haldane, L. Balents, Phys. Rev. Lett. {\bf 99}, 196802 (2007).
\bibitem{streda}
P. St\v{r}eda, J. Phys. C: Solid State Phys. {\bf 15}, L717 (1982).
\\
\end{thebibliography}
\end{document}